**Correlating the nanostructure of Al-oxide with deposition conditions and dielectric contributions of two-level systems in perspective of superconducting quantum circuits**

S. Fritz[1,*], A. Seiler[2], L. Radtke[2], R. Schneider[1], M. Weides[2], G. Weiß[2] and D. Gerthsen[1,*]

[1]Laboratory for Electron Microscopy, Karlsruhe Institute of Technology (KIT),

76131 Karlsruhe, Germany

[2]Physikalisches Institut, Karlsruhe Institute of Technology (KIT), 76131 Karlsruhe, Germany

[*]Correspondence and requests for materials should be addressed to D.G. (email: dagmar.gerthsen@kit.edu) or S.F. (email: stefan.fritz@kit.edu)

**Abstract**

This work is concerned with Al/Al-oxide(AlO$_x$)/Al-layer systems which are important for Josephson-junction-based superconducting devices such as quantum bits. The device performance is limited by noise, which has been to a large degree assigned to the presence and properties of two-level tunneling systems in the amorphous AlO$_x$ tunnel barrier. The study is focused on the correlation of the fabrication conditions, nanostructural and nanochemical properties and the occurrence of two-level tunneling systems with particular emphasis on the AlO$_x$-layer. Electron-beam evaporation with two different processes and sputter deposition were used for structure fabrication, and the effect of illumination by ultraviolet light during Al-oxide formation is elucidated. Characterization was performed by analytical transmission electron microscopy and low-temperature dielectric measurements. We show that the fabrication conditions have a strong impact on the nanostructural and nanochemical properties of the layer systems and the properties of two-level tunneling systems. Based on the understanding of the observed structural characteristics, routes are derived towards the fabrication of Al/AlO$_x$/Al-layers systems with improved properties.

**Introduction**

Intense efforts have been devoted to possible realizations of quantum information processing during the past two decades. One of the technologically most advanced approaches at present relies on superconducting circuits[1]. Their basic units, quantum bits (qubits), contain Josephson junctions (JJs) which are worldwide mainly fabricated on the basis of Al/AlO$_x$/Al-layer systems. Complex superconducting circuits will contain a large number of qubits and JJs whose properties, in particular coherence and relaxation times, and homogeneity will be crucial for further progress of the technology. Coherence times up to $10^{-3}$ s have been achieved[2] by the elimination of various sources of noise in well-shielded single-qubit experiments but further improvements are necessary for complex quantum circuits. This has led to intense investigations of sources of noise, such as flux noise[3], fluctuations of the critical current[4], charge noise induced by surface and gap states[5], and contaminations at interfaces[6]. A major source of noise and decoherence in qubits is assigned to two-level defects, commonly denoted as two-level systems (TLS). Several theoretical and some experimental studies on TLS properties resulted in suggestions regarding possible sources of TLS[6-10] but, despite of these efforts, the microscopic origin of TLS is still under debate.

The properties of individual TLS were investigated in only few experimental studies. Using qubits, resonance techniques were applied to spectroscopically identify TLS in the microwave frequency regime (few GHz) which are characterized by electric and elastic dipole moments[7,8]. Recent theoretical studies focused on the intrinsic properties of the AlO$_x$-tunnel barrier, where bistable defects, e.g. delocalized oxygen atoms, could be a source of TLS[9,10]. It has been indeed well known for decades that two-level atomic tunneling systems dominate the low-temperature properties of glasses and other disordered solids[11,12]. Small groups of atoms, the nature of which is not further specified, are assumed to tunnel between two energetically almost equivalent configurations. The TLS are modeled as particles in double-well potentials with a broad distribution of the relevant parameters. The resulting density of states is essentially constant, though material-specific, causing a logarithmic dependence of the dielectric permittivity as a function of temperature[13,14]. For the case of a disordered $AlO_x$-film, this logarithmic dependence scales with the combined effect of TLS density of states per volume and energy and the squared average TLS dipole moment[15].

Numerous studies are concerned with the fabrication and properties of $AlO_x$-based tunnel barriers. The standard procedure involves thermal or plasma-assisted surface oxidation of an Al-layer[16–20], typically performed at room temperature. This is a self-limiting process and results in an amorphous $AlO_x$-layer with a thickness of 1-3 nm. Optimization of the tunnel barrier can be achieved by studying, e.g., the critical current density of JJs as a function of oxidation pressure and oxidation time[21]. A variety of different techniques were used in the past to study JJ properties but only few dedicated analytical transmission electron microscopy (TEM) investigations were undertaken to analyze the structural and chemical properties of amorphous $AlO_x$-tunnel barriers on the nanoscale[22–25]. A recent study by Zeng et al.[26] focuses on the analysis of locally acquired pair distribution functions and concludes that the $AlO_x$-tunnel barrier is O-deficient. These results may give hints on the microscopic nature of TLS. We note that other applications of $Al/AlO_x/Al$- and $Nb/AlO_x/Nb$-based JJs such as X-ray detectors[27], voltage standards[28] and superconducting quantum devices[29] will also profit from optimized JJs because their properties also sensitively depend on noise.

To make further progress towards the reduction of noise induced by TLS, in this work the nanochemical and nanostructural properties of $Al/AlO_x/Al$-layer systems are correlated with deposition conditions and low-temperature capacitance measurements to assess the TLS contribution to the dielectric permittivity. TEM and scanning transmission electron microscopy (STEM) combined with electron energy loss spectroscopy (EELS) were applied to analyze the nanochemistry and nanostructure of four differently fabricated $Al/AlO_x/Al$-layer systems.

$AlO_x$-layers with a thickness of 20 to 30 nm are necessary to obtain a sufficiently high electric resistance and small electric field strength for capacitance measurements. We find pronounced differences of the dielectric TLS contribution and structural and chemical properties of the analyzed $Al/AlO_x/Al$-layer systems, which suggest measures on how to improve the properties of the $AlO_x$-tunnel barriers in JJs in the future.

**Experimental Results**

Four differently produced $Al/AlO_x/Al$-layer systems were investigated. The first sample was fabricated in a *MEB 550S (PLASSYS Bestek, Marolles-en-Hurepoix, FR)* electron-beam deposition system. After deposition of an Al-layer, dynamic oxidation with pure $O_2$ at room temperature was applied to form the $AlO_x$-layer. The same system was used to fabricate a second sample, where $AlO_x$-formation took place under UV-illumination to enhance dynamic oxidation. These two samples are denoted as EBPlas and EBPlas-UV. $AlO_x$-layer thicknesses between 20 and 30 nm were obtained by repeating Al-deposition and dynamic oxidation several times to achieve a sufficiently high electric resistance for the capacitance measurements (see Methods). The third sample, denoted as EBLes, was fabricated in a different electron-beam deposition system (*PVD 75, Kurt J. Lesker Company, Hastings, UK*). Electron-beam deposition was applied not only for the Al-layers but also for the $AlO_x$-layer using stoichiometric $Al_2O_3$-pellets. The fourth sample SPUT was fabricated in a home-built sputter deposition system. Al-deposition was performed using an Al-target in an Ar-plasma, while the $AlO_x$-film was deposited from the same Al-target in a reactive Ar/O-plasma.

Fig. 1 shows overview bright-field (BF) TEM cross-section images of all samples. Grains with a large average lateral size of $(196 \pm 89)$ nm are observed in the lower Al-layer of EBPlas (Fig. 1a). The rather homogenous layer thickness leads to a small corrugation of the lower $Al/AlO_x$-interface. The thickness of the $AlO_x$-layer varies considerably and increases the corrugation at the upper Al/AlO$_x$-interface. The most pronounced thickness variations and distortions of the AlO$_x$-layer are found in regions where Al-grain boundaries intersect the Al/AlO$_x$-interface (white arrows in Fig. 1a). Thickness homogeneity of the AlO$_x$-layer and interface corrugation are improved in EBPlas-UV (Fig. 1b). The average lateral grain size (189 ± 90) nm in the lower Al-layer is similar as in EBPlas. The microstructure of EBLes (Fig. 1c) differs considerably from EBPlas and EBPlas-UV although it was also fabricated in an electron-beam deposition system. The Al-layers are substantially corrugated and consist of small Al-grains with sizes of (54 ± 23) nm. The AlO$_x$-layer varies strongly in thickness.

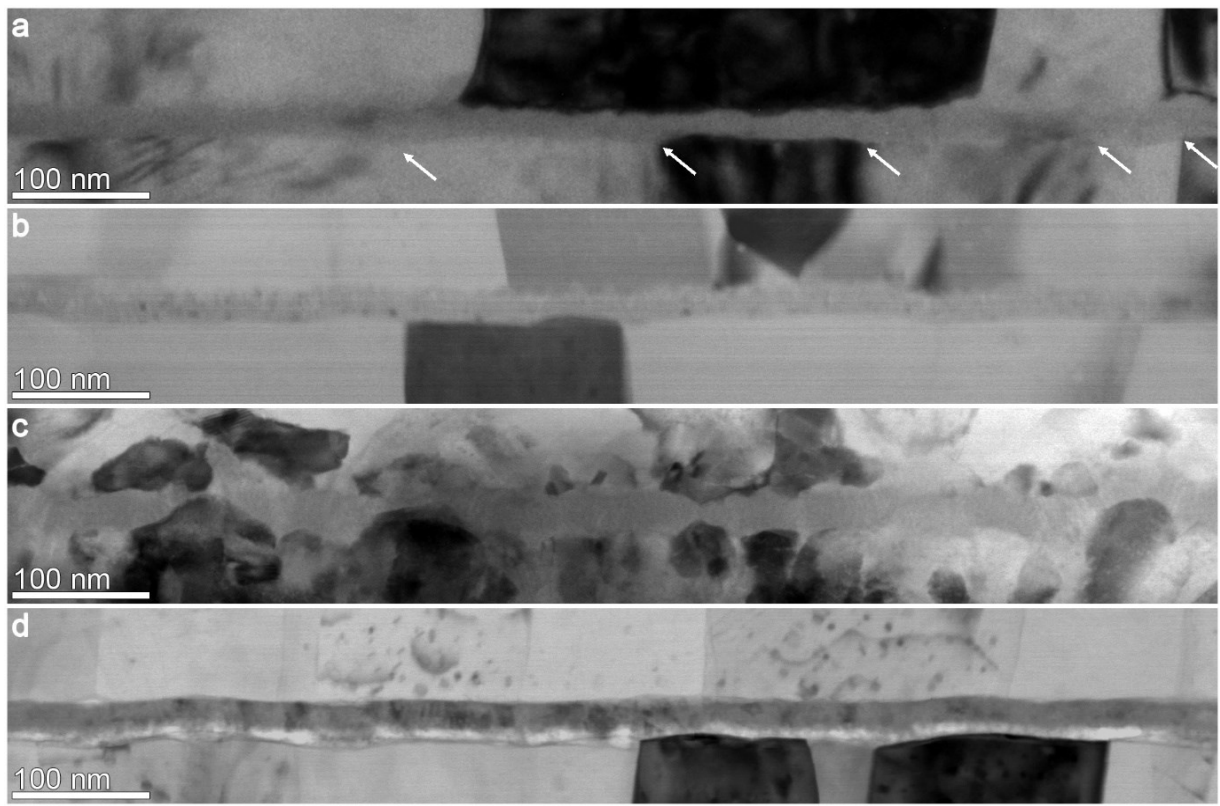

**Figure 1**. Structure of Al/AlO$_x$/Al-layer systems fabricated by different techniques. Overview BF TEM cross-section images of (a) EBPlas, (b) EBPlas-UV (c) EBLes and (d) SPUT. Grain boundaries in the bottom Al-layer of EBPlas (a) are marked by white arrows.

The forth sample SPUT (Fig. 1d) contains Al-layers with an average lateral grain size of (117 ± 62) nm and an AlO$_x$-layer with the most homogeneous thickness of all samples. However, the AlO$_x$-layer appears to be subdivided into a lower sublayer with bright contrast and a darker upper part.

High-resolution TEM (HRTEM) images in Fig. 2 reveal further structural details of the $AlO_x$-layers. EBPlas and EBPlas-UV contain crystalline regions of a few nm size embedded in the amorphous $AlO_x$-matrix (nanocrystals marked by circles in Fig. 2a,b). The crystal structure of the nanocrystals was determined by comparing the two-dimensional Fourier transformation (FT) pattern of these regions with calculated diffraction patterns. The FT patterns of all analyzed nanocrystals in EBPlas and EBPlas-UV (cf. representative examples in Fig. 2a,b) agree with calculated diffraction patterns of Al (face-centered cubic structure, space group Fm-3m, lattice parameter a = 4.06 Å[30]). An almost entirely crystalline $AlO_x$-layer is found in EBLes (Fig. 2c) where electron-beam deposition from a stoichiometric $Al_2O_3$-target was used for $AlO_x$-layer deposition. FT analysis (inset in Fig. 2c) shows that the $AlO_x$ crystallizes in the $\gamma$-$Al_2O_3$ phase with a defect cubic spinel structure (space group Fd-3m, lattice parameter a = 7.91 Å[31]). The $AlO_x$-layer of the sample SPUT (Fig. 2d) is subdivided into an amorphous lower part with an average thickness of 8.6 ± 2.1 nm and a crystalline upper part, which consists of $\gamma$-$Al_2O_3$.

STEM-EELS was applied to obtain information on the local chemical composition and bonding characteristics of Al and O within the $AlO_x$-layers. The energy loss near edge structure (ELNES) of the Al-$L_{2,3}$ and O-K edges is shown in the EELS spectra (Fig. 3a-c) for all four samples. Spectra of a crystalline $\gamma$-$Al_2O_3$ reference sample obtained from a nanoparticle (nanoparticulate powder *by Carl Roth GmbH + Co KG, Karlsruhe, GER*) are included for comparison.

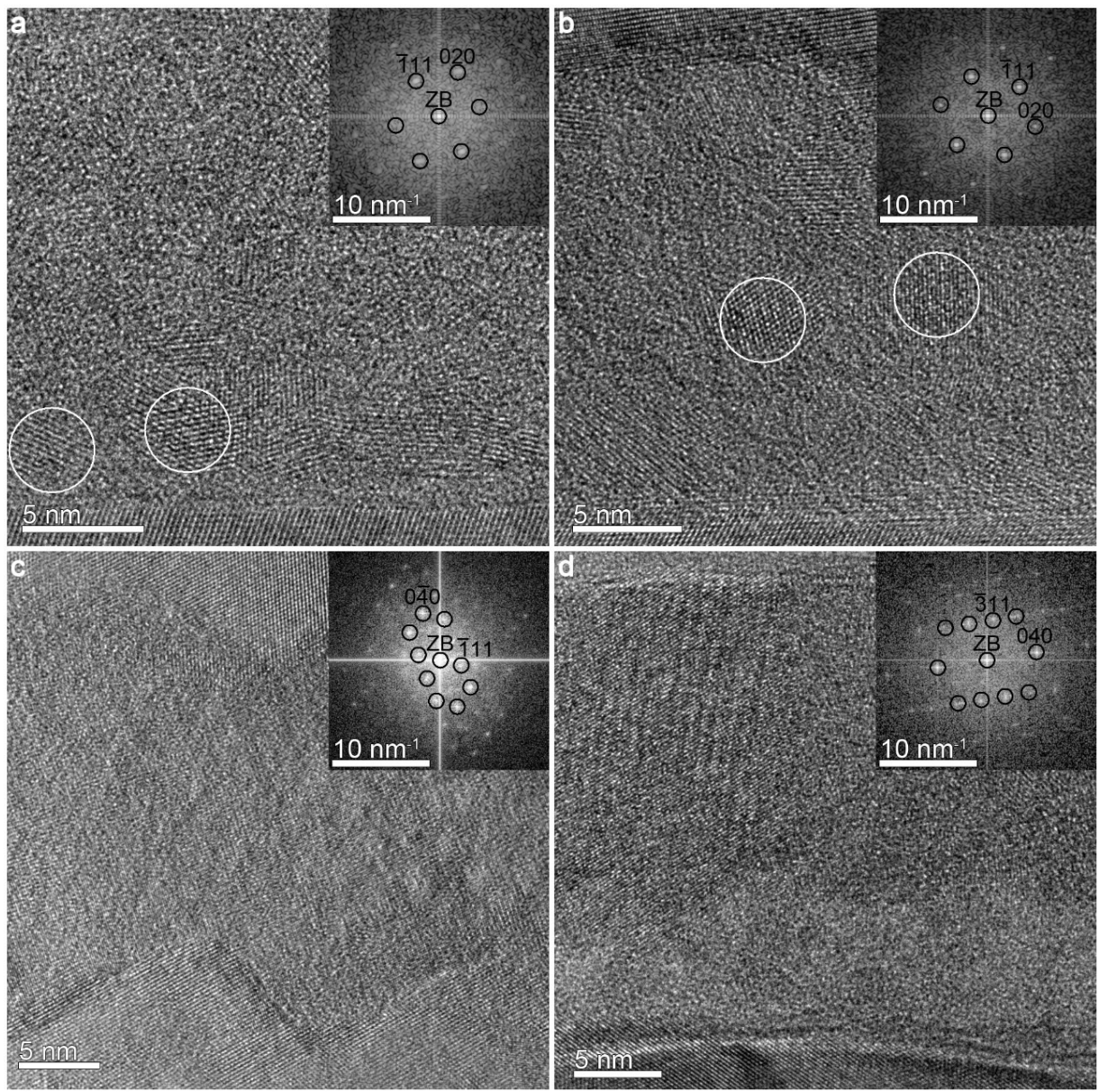

**Figure 2.** Structure of the AlO$_x$-layers from differently fabricated Al/AlO$_x$/Al-layer systems. High-resolution cross-section TEM images of (a) EBPlas with FT pattern of an Al-nanocrystal in [110] zone-axis, (b) EBPlas-UV with FT pattern of an Al-nanocrystal in [110] zone-axis, (c) EBLes with FT pattern of γ-Al$_2$O$_3$ in the [101] zone-axis and (d) SPUT with FT pattern of γ-Al$_2$O$_3$ in the [103] zone-axis.

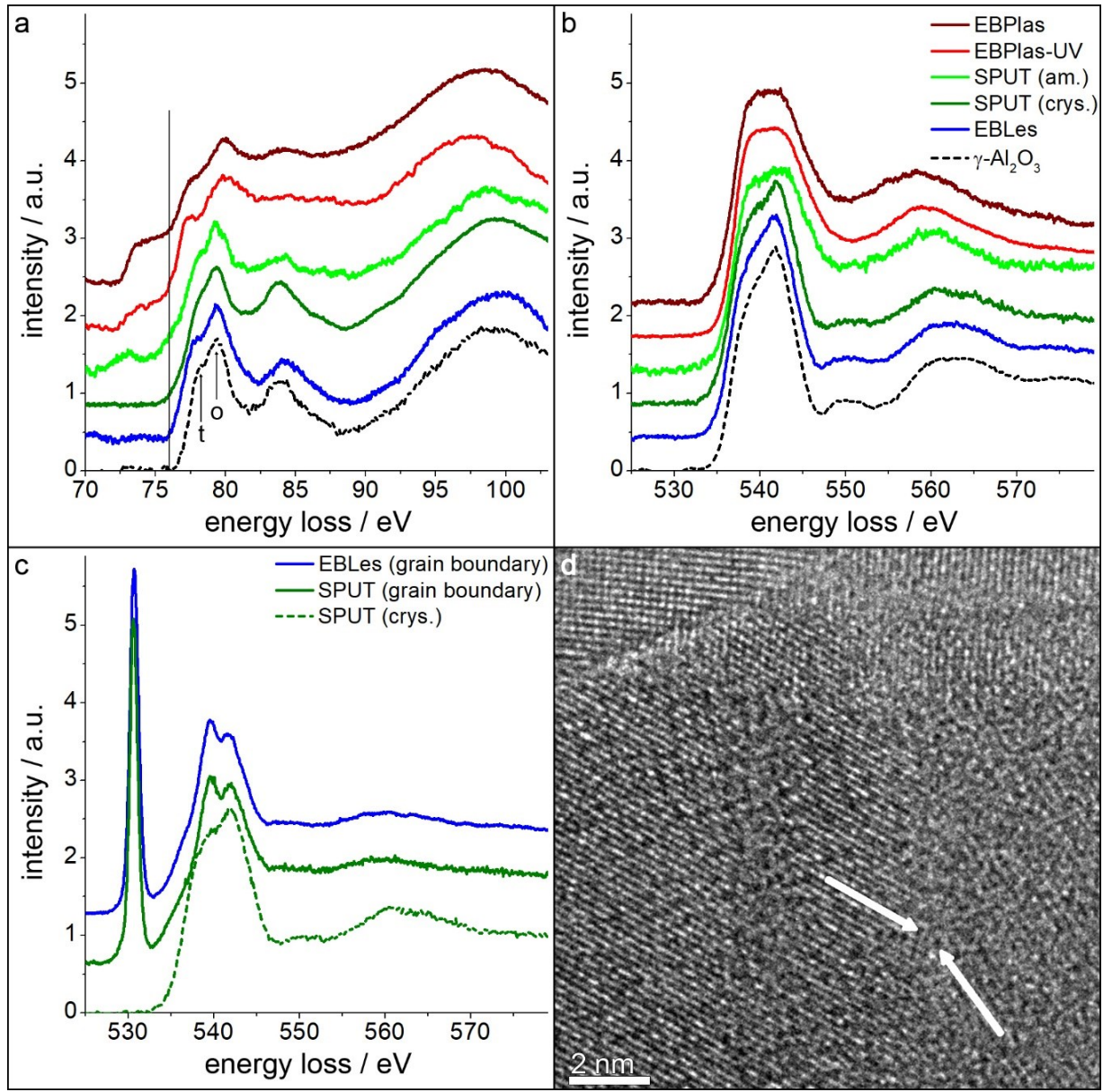

**Figure 3.** EELS spectra showing the ELNES of the Al-$L_{2,3}$ and O-K edges in the AlO$_x$-layers of all samples and a γ-Al$_2$O$_3$ reference specimen. Spectra of (a) Al-$L_{2,3}$ and (b) O-K edges acquired in amorphous (EBPlas, EBPlas-UV, SPUT) and crystalline (EBLes and SPUT) regions. A spectrum of a crystalline γ-Al$_2$O$_3$ reference specimen is included. The edge onset of γ-Al$_2$O$_3$ at 76 eV is marked by a black line in (a). 't' and 'o' correspond to tetrahedrally and octahedrally coordinated Al-atoms. (c) shows the O-K edge acquired at AlO$_x$-grain boundaries of EBLes and SPUT (solid lines) and within a crystalline region (dashed line) with (d) corresponding HRTEM image of a grain boundary region in SPUT. A boundary between two Al$_2$O$_3$-grains is marked by white arrows.

The Al-$L_{2,3}$ edge of γ-Al$_2$O$_3$ starts at 76 eV (Fig. 3a, black line). The arrows mark two signals at energy losses of 77.9 eV (arrow labeled 't') and 79.4 eV (arrow labeled 'o') which can be associated with tetrahedrally and octahedrally coordinated Al-atoms[32,33] as expected in γ-Al$_2$O$_3$.

The Al-$L_{2,3}$ edge of EBLes and the crystalline AlO$_x$-region of SPUT (blue and green lines in Fig. 3a) agree well with the ELNES of γ-Al$_2$O$_3$. The ELNES of amorphous AlO$_x$ in EBPlas and EBPlas-UV (brown and red line in Fig. 3a) looks distinctly different. The onset is shifted to 72.5 eV which indicates that metallic Al is also present in the analyzed region[33]. The pronounced maximum at 84 eV in γ-Al$_2$O$_3$ is not present anymore in amorphous AlO$_x$ due to the lack of a medium-range ordered structure[34]. The first peak after the edge onset contains signatures of tetrahedrally and octahedrally coordinated Al-atoms but with slightly shifted energies. These changes can be attributed to distorted bonds (variations of bond lengths and angles) in the amorphous material. Moreover, Zheng at al.[26] have shown that 2-, 3- and 5-fold coordinated Al-atoms are also present in amorphous AlO$_x$ which are expected to influence the Al-$L_{2,3}$ ELNES. Fig. 3b presents spectra of the O-K edge of all samples including the γ-Al$_2$O$_3$ reference sample. The ELNES features of EBLes and in the crystalline AlO$_x$-region of SPUT agree well with the O-K edge of γ-Al$_2$O$_3$. It is characterized by an intense peak at the edge onset between 533 and 534 eV and two maxima at energy losses of 550 and 563 eV. The O-K edge in the amorphous AlO$_x$-layer in EBPlas and EBPlas-UV is different and shows a broadened first peak followed by only one maximum at an energy loss of 557 eV.

Fig. 3c shows again spectra of the O-K edge acquired in EBLes and SPUT. It contains two spectra that were taken with the electron beam positioned at boundaries between crystalline Al$_2$O$_3$-grains as indicated by the white arrows in the HRTEM image of SPUT (cf. Fig. 3d). A dramatic change of the ELNES of the O-K edge is observed at such positions (spectra plotted with solid lines in Fig. 3c) compared to the O-K edge inside crystalline γ-Al$_2$O$_3$ grains (spectrum with dashed line in Fig. 3c). A sharp first peak at 530.7 eV shows up at grain boundaries and a distinct peak splitting with maxima at 539.6 eV and 541.9 eV is observed. The sharp features of the O-K edge at grain boundaries indicate that molecular oxygen (O$_2$) is the origin of the specific ELNES in this case. This assumption is confirmed by literature data of XAS and EELS spectra of $O_2$[35,36]. In X-ray absorption spectroscopy of $O_2$, the same features are observed, which correspond to transitions from the occupied 1s to unoccupied $\pi^*$ (sharp line at 530.5 eV) and $\sigma^*$ states (539.5 and 542 eV)[37]. We like to emphasize at this point that $O_2$ is inherently present at grain boundaries and is not an effect of electron-beam damage because the sharp spectral features do not develop in the course of electron-beam illumination and do not change even during extended observation in the electron microscope.

| sample | region | chemical composition $AlO_x$ |
|---|---|---|
| EBPlas | amorphous | $0.48 \pm 0.04$ |
| | crystalline | Al |
| EBPlas-UV | amorphous | $1.10 \pm 0.06$ |
| | crystalline | Al |
| EBLes | amorphous | $1.31 \pm 0.03$ |
| | crystalline | $1.50 \pm 0.04$ |
| | grain boundaries | $1.73 \pm 0.10$ |
| SPUT | amorphous | $1.18 \pm 0.05$ |
| | crystalline | $1.49 \pm 0.03$ |
| | grain boundaries | $1.82 \pm 0.10$ |

**Table 1**. Chemical composition in different regions of the $AlO_x$-layers in the investigated samples.

The composition of the $AlO_x$-layers was quantitatively analyzed by evaluation of the intensities of the Al-$L_{2,3}$ and O-K edges (cf. Table 1). A high O-deficiency with a chemical composition of $AlO_{0.5}$ is found in EBPlas. The O-content raises to $AlO_{1.1}$ in EBPlas-UV due to UV-enhanced oxidation. Small amorphous regions in EBLes consist of $AlO_{1.3}$, which is similar to $AlO_{1.2}$ in the lower amorphous part of the $AlO_x$-layer in SPUT. Stoichiometric $AlO_{1.5}$ is formed in the crystalline regions of EBLes and SPUT. O-excess is present at boundaries between crystalline grains of SPUT ($AlO_{1.8}$) and EBLes ($AlO_{1.7}$) which is consistent with the observation of molecular $O_2$ in EELS spectra (Fig. 3c). Overall, amorphous $AlO_x$ exhibits an O-deficiency, which depends on deposition technique and fabrication conditions. Crystalline regions consist of pure Al or stoichiometric γ-Al$_2$O$_3$. Grain boundaries in crystalline γ-Al$_2$O$_3$ show an O-excess due to the presence of molecular O$_2$.

To investigate the impact of these different microstructures on the low-energy excitations in the material, dielectric measurements at low temperatures were performed. Numerous previous studies of amorphous solids at low temperatures[11,12,38] have demonstrated that capacitance variations allow to derive the dielectric TLS contribution to $\epsilon$, which is given by

$$\epsilon = \epsilon_0 \cdot \epsilon_r = \epsilon_0(1 + \chi_{Al_2O_3} + \chi_{TLS}) \tag{1}$$

where $\chi_{Al_2O_3}$ describes the temperature-independent susceptibility for the Al$_2$O$_3$-background. The temperature-dependent susceptibility $\chi_{TLS}(T)$, given by [14]

$$\chi_{TLS}(T) = \kappa \cdot \log\left(\frac{T_0}{T}\right) \tag{2}$$

is attributed to TLS at low temperatures. The prefactor $\kappa \propto N \cdot p^2$ combines the TLS density of states per volume and energy $N$ and the squared average TLS dipole moment $p^2$ [15]. The TLS contribution to the dielectric permittivity $\epsilon$ of the AlO$_x$-layer can be derived from capacitance measurements at low temperature for a constant measurement frequency and a constant external electric field. The capacitance variation $\Delta C = C(T) - C_{T_0}$ is measured with respect to reference capacity $C_{T_0}$ at an arbitrary temperature $T_0$ and leads to

$$\Delta C/C_{T_0} = \Delta\epsilon/\epsilon_{T_0} = \Delta\chi_{TLS}/\epsilon_{T_0} \tag{3}$$

as $\chi_{Al_2O_3}$ is expected to be constant and therefore $\Delta\chi_{Al_2O_3} = 0$.

Fig. 4 shows the variation of the dielectric permittivity $\Delta\epsilon$ as a function of the logarithm of the temperature. $\Delta\epsilon$ is normalized with $\epsilon_{T_0}$ where $T_0$ was chosen to be 200 mK for SPUT and EBPlas-UV. The capacitance measurements were performed at 1 kHz and an applied voltage of 1 mV. The dielectric response of the disordered AlO$_x$-layers shows signatures similar to glasses which allows to extract the underlying TLS properties and their dependence on the fabrication method. Starting from the lowest temperature the dielectric permittivity decreases with rising temperature, because TLS with energy splitting $E$ are thermally excited when $k_B T$ becomes larger than $E$. Therefore TLS cannot be polarized by the applied electric field. Due to the broad distribution of TLS energies, this decrease depends on the logarithm of the temperature and is proportional to $-\kappa$ according to Eq. (2). A minimum of $\Delta\epsilon$ appears when the relaxation rate of the dominating TLS with $E \approx k_B T$ becomes equal to the measuring frequency. At still higher temperatures the energy relaxation rate of TLS exceeds the measurement frequency and leads to an increase of the permittivity. At temperatures $T_{min}$, where the permittivity minimum occurs, the TLS relaxation is dominated by one-phonon processes. $T_{min}$ gives therefore a measure of the coupling strength of TLS to phonons – the smaller $T_{min}$, the stronger the coupling.

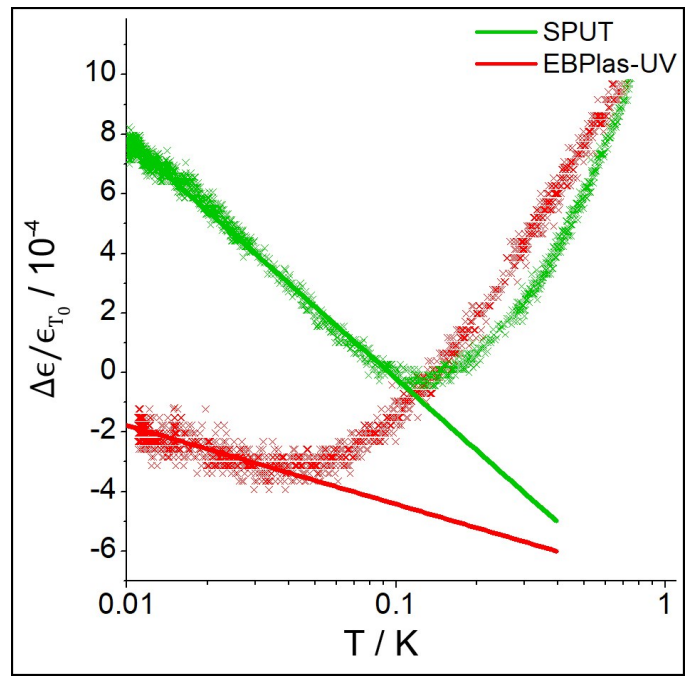

**Figure 4.** Low-temperature dielectric capacitance measurements of EBPlas-UV and SPUT. $\Delta\epsilon/\epsilon_{T_0}$ is plotted as a function of the logarithm of the temperature. The solid lines are linearly fitted to the data and represent the slope of $\Delta\epsilon/\epsilon_{T_0}$ at low temperatures.

The temperature dependences of the dielectric permittivity of SPUT and EBPlas-UV are remarkably different (cf. Fig. 4) in two aspects. EBPlas-UV shows a much weaker decrease of the dielectric permittivity than SPUT indicating that $\kappa$ is reduced in EBPlas-UV. In addition, the TLS relaxation rate for EBPlas-UV exceeds the frequency of the applied ac-electric field of 1 kHz at 50 mK – a behavior that is known from many other disordered solids. However, the onset of relaxation in SPUT at 150 mK indicates an unusual weak coupling to phonons. This difference in coupling strength to phonons between SPUT and EBPlas-UV can be most likely attributed to different microscopic origins of the relevant TLS.

For EBPlas the TLS contributions to the dielectric permittivity could not be determined due to a high dc-loss obscuring the signal and making it impossible to measure small capacitance variations. This could be the result of an increased conductivity due to the low O-content ($AlO_{0.5}$) in this layer. The dc-loss can also be caused by problems with the epoxy used for contacting the capacitor, which reacts with the Al-layers. Failure of lithography prevented sample preparation for capacitance measurements of EBLes.

**Discussion**

Analytical TEM shows that the nanostructural and nanochemical properties of $AlO_x$-layers in $Al/AlO_x/Al$-layer systems vary considerably depending on the deposition system and conditions. In the following we will first discuss the properties of the lower Al-layer and their effect on the $AlO_x$-layers. We then focus on the properties of the $AlO_x$-layers and finally discuss the correlation of the measured dielectric TLS contribution $\kappa$ of samples EBPlas-UV and SPUT with their nanoscale structural and chemical properties.

*Properties of the lower Al-layer*

Fig. 1a-d demonstrate that the morphology of the $AlO_x$-layers is strongly influenced by the lower Al-layer and motivate efforts to optimize its properties. $AlO_x$-thickness variations strongly affect the local tunnel current, which scales as $e^{-t/\lambda}$ with the local barrier thickness $t$ and the attenuation length $\lambda$. Zheng et al.[24] have already shown that less than 10 % of the total

barrier area is active in the tunneling process due to thickness fluctuations in the tunnel barriers of JJs studied in their work.

Pronounced $AlO_x$-thickness variations occur at grain boundaries of the underlying Al-layer (cf. regions indicated by arrows Fig. 1a) or on bottom Al-layers with strongly corrugated surface topography (cf. Fig. 1c). $AlO_x$-layer sections with homogeneous thickness are observed on top of single Al-grains with planar surface. There are two possible origins for the thickness increase of $AlO_x$ at Al-grain boundaries. One effect is grain boundary grooving due to the establishment of a local mechanical equilibrium between Al-surface tension and grain boundary energy. The effect was originally studied by Mullins[39] and recently discussed with respect to $AlO_x$-layers in JJs by Nik et al.[40]. Another possibility can be enhanced oxygen diffusion along grain boundaries[41]. Irrespective of the mechanism, large lateral grain sizes with homogeneous thickness are desirable to avoid fluctuations of the $AlO_x$-thickness and tunnel current caused by it. Nik et al.[40] observed a log-normal grain size distribution and found the average lateral grain sizes to be approximately by two to three times larger than the Al-layer thickness. These characteristics are typical for normal grain growth[42], which is also observed in our samples EBPlas, EBPlas-UV and SPUT where a log-normal grain size distribution and lateral average grain sizes of twice the Al-film thickness were found.

The microstructure of the Al-layer in EBLes clearly deviates from the behavior of the other samples because the average grain size of 54 nm corresponds to only 50 % of the film thickness and the surface is strongly corrugated. This is on first sight surprising because EBLes was also fabricated by electron-beam deposition as EBPlas and EBPlas-UV (although in a different system) with similar nominal deposition rates (0.2 nm/s for EBPlas and EBPlas-UV vs. 0.13 nm/s for EBLes) without intentional cooling or heating of the substrate. We attribute the observed differences to the effect of different $O_2$-partial pressures during Al-deposition as previously reported for electron-beam deposited Al-films by Verkerk et al.[43]. Al-oxidation can take place at higher $O_2$-partial pressures during Al-deposition, which then initiates the formation of secondary Al-grains on the oxidized Al-surface. Nucleation of new Al-grains within the Al-film is clearly recognized in Fig. 1c and contributes to the reduction of the average grain size. Oxidation may indeed have occurred during Al-deposition in the Lesker electron-beam deposition system where the (overall) pressure of $1.5 \cdot 10^{-6}$ mbar was one order of magnitude higher than in the Plassys system ($1.5 \cdot 10^{-7}$ mbar). Oxidation may have been also favored by the smaller deposition rate of 0.13 nm/s in the Lesker system compared to 0.2 nm/s in the Plassys system. Another origin of the reduced average grain size may be found in the substrate properties. In addition to the native $SiO_x$-layer, a 10-13 nm thick carbon contamination layer was present between bottom Al-layer and substrate for EBLes, despite identical substrate pretreatment. The larger substrate roughness provides a higher density of grain nucleation sites and contributes to the observed small average grain size.

*Properties of the Al-oxide layers*

The composition of the $AlO_x$-layer tunnel barrier in JJs could be a decisive property for the TLS density as suggested by theoretical studies[12]. The measured compositions (cf. Table 1) reveal substantial O-deficiencies depending on the oxidation procedure. All amorphous $AlO_x$-layers are sub-stoichiometric in contrast to $\gamma$-$Al_2O_3$ in EBLes and SPUT, which contain the expected O-concentration. Dynamic oxidation with $O_2$ with 10 sccm for 12.5 min for EBPlas yields strongly O-deficient amorphous $AlO_{0.5}$. The O-concentration increases considerably to $AlO_{1.1}$ if UV-illumination is used (EBPlas-UV) in combination with a slightly increased $O_2$-flux of 12.5 sscm. The interaction of UV-photons and $O_2$ enhances the dissociation rate of $O_2$ at the surface and additionally creates energetically activated O-ions which help to reduce the activation barrier for chemisorption[44]. The resulting negatively charged O-ions diffuse into the $AlO_x$-layer and bind to positively charged Al-ions. This effect can be further enhanced by low-energy electron-bombardment of the surface to drive the charged O-ions deeper into the $AlO_x$-layer[20].

Monocrystalline Al-inclusions with a few nm size are embedded in the amorphous $AlO_x$-layer of EBPlas and EBPlas-UV (cf. Fig. 2a,b). They are only present in samples fabricated in the Plassys system where repeated Al-deposition/dynamic-oxidation (cf. Methods) were applied to obtain $AlO_x$-layers with sufficient thickness and high resistance. TEM studies of samples obtained with a single dynamic oxidation process (not presented here) show that the $AlO_x$-layer has a maximum thickness of about 3 nm. To obtain thicker $AlO_x$-layers, approximately 1 nm Al is deposited after the first oxidation step and dynamically oxidized subsequently. This process step is repeated up to ten times resulting in a ~20 nm thick $AlO_x$-layer. We assume that a 1 nm Al-deposition does not lead to a homogenous Al-coverage but to the formation of Al-islands, which are not fully oxidized in the subsequent oxidation process leading to embedded Al-inclusions.

The amorphous lower part of SPUT with a composition of $AlO_{1.2}$ contains a rather high O-concentration. The enhancement of O-concentration compared to dynamic oxidation is attributed to the plasma-assisted deposition process where several factors contribute to the high O-content. First, the Ar/O-plasma generates dissociated and more reactive O-ions at the $AlO_x$-surface. Second, a non-negligible concentration of the sputtered Al-atoms could be already oxidized in the gas phase and, third, the surface of the Al-sputter target may be partially oxidized during the sputter process.

An interesting observation is the transition from amorphous $AlO_{1.2}$ to crystalline $\gamma$-$Al_2O_3$ (cf. Fig. 2c) on a substrate that was not intentionally heated during the deposition process. A clue to understand this phenomenon is provided the work by Jeurgens *at al.*[45]. They considered the thermodynamic stability of amorphous and crystalline $Al_2O_3$ on Al-surfaces with different crystallographic orientation. By considering the total Gibbs free energy of amorphous and crystalline $Al_2O_3$-films they found amorphous $Al_2O_3$ to be more stable than γ-$Al_2O_3$ up to a critical thickness that depends on the crystallographic orientation of the Al-surface and the temperature. Typical calculated values for the critical thickness at room temperature are 2 nm for Al(100) and 4 nm for Al(110) which are in reasonable agreement with the measured thickness of 8.6 ± 2.1 nm taking into account that the polycrystalline Al-film provides a variety of different, not necessarily low-index surfaces.

Interestingly, EBLes contains an almost completely nanocrystalline γ-$Al_2O_3$ layer with only small amorphous regions which was obtained without intentional substrate heating. We attribute its formation to the comparatively high pressure of $10^{-5}$ mbar during oxidation in Lesker system (higher base pressure and evaporation from $Al_2O_3$-pellets) and the very low evaporation rate of 0.03-0.04 nm/s. This observation underlines that high temperatures are not required for γ-$Al_2O_3$ formation. An extremely slow growth rate in an atmosphere with a sufficient $O_2$-concentration facilitates uptake of a rather large concentration of oxygen that is favorable for the formation of crystalline γ-$Al_2O_3$. Although the formation of a crystalline $Al_2O_3$-layer is on first sight desirable, the discussion in the next section will show that $O_2$ embedded in grain boundaries has a detrimental effect on $\kappa$ as shown by the different dielectric behavior of EBPlas-UV and SPUT (cf. Fig. 4).

*Correlation of dielectric TLS contribution and Al-oxide nanostructure*

In the following we discuss the correlation between nanoscale structural and chemical properties and dielectric permittivity for EBPlas-UV and SPUT. With respect to the possible origin of TLS, DuBois *et al.*[12] suggested delocalized O-atoms in amorphous $AlO_x$ as possible sources of TLS where O-atoms in an Al-O-Al chain can occupy different positions and induce this way a TLS and charge dipole. Another indication that oxygen is connected with TLS was provided by the study of Tan *et al.*[20] who detected weakly bound O-atoms by X-ray photoelectron studies and suggested that these O-atoms could be metastable defects and contribute to $1/f$ noise in JJs. These previous studies suggest to focus on the O-content and O-bonding to understand the strongly different dielectric permittivity of SPUT and EBPlas-UV. SPUT contains a complex $AlO_x$-layer where the upper part consists of nanocrystalline $\gamma$-$Al_2O_3$ with a considerable content of $O_2$ located at grain boundaries (cf. Table 1 and Fig. 3c). The lower part with an average thickness of 8.6 nm (about 40 % of the overall thickness) is amorphous with a composition of $AlO_{1.2}$. The $AlO_x$-layer in EBPlas-UV is amorphous with some embedded nanocrystalline Al-inclusions. The O-concentrations of SPUT ($AlO_{1.2}$) and EBPlas-UV ($AlO_{1.1}$) agree within the error limits and suggest comparable amorphous $AlO_x$-properties. Assuming that the O-content is correlated with the defect concentration that determines the TLS density, we expect similar values $\kappa$ values for both samples. However, the dielectric measurements (Fig. 4) show that $\kappa$ is much larger in SPUT than in EBLes-UV. This suggests the crystalline part of the Al-oxide layer in SPUT to be responsible for the higher value $\kappa$. Thus, it either contains a higher TLS density of states in the crystalline layer or the individual TLS must have a larger electric dipole moment. While crystalline $Al_2O_3$ has shown beneficial effects if present as epitaxial layer (80 % reduced TLS density[46] and reduced TLS coupling strength[47]), the nanocrystalline $\gamma$-$Al_2O_3$ contains a high density of grain boundaries with embedded $O_2$. We tentatively interpret the high $\kappa$ to be induced by loosely bound oxygen that forms atomic tunneling systems. As the number of O-atoms at grain boundaries should be much smaller than the number of defects in an amorphous layer, it is likely that TLS located at $\gamma$-$Al_2O_3$ grain boundaries have a large dipole moment due to the high electronegativity of the oxygen atoms. The observation of the unusual high temperature of 150 mK for the onset of relaxation processes in the dielectric measurements supports this interpretation because such loosely bound O-atoms may couple rather weakly to lattice vibrations (cf. Fig. 4). This interpretation is in line with recent work by Kumar et al.[48] who found $O_2$ on the surface of Al/$AlO_x$/Al-based qubits to be the dominant source of $1/f$ noise in their structures.

**Conclusions**

Analytical electron microscopy studies and low-temperature capacitance measurements have demonstrated the strong impact of the fabrication conditions on the properties of Al/AlO$_x$/Al-layer systems in this work. The main results and conclusions with respect to optimization of the Al/AlO$_x$/Al-structures for JJs are:

- The morphology of the AlO$_x$-layer is predominantly determined by the structure and morphology of the bottom Al-layer and even the surface morphology of the underlying substrate plays a role. Grain boundaries intersecting the lower AlO$_x$/Al-interface lead to thickness variations of the AlO$_x$-layer. AlO$_x$-layers with homogeneous thickness can be expected from a flat and, ideally, epitaxial bottom Al-electrode layer without grain boundaries.

- The O-content in amorphous AlO$_x$ varies strongly depending on the deposition conditions. The highest O-contents were obtained by UV-assisted dynamic oxidation of Al and sputter-deposition in an Ar/O-plasma. Since the TLS density is likely to be correlated with the defect density, the O-content in amorphous AlO$_x$ should be as close as possible to AlO$_{1.5}$ to minimize the concentration of low-coordinated atoms.

- Nanocrystalline γ-Al$_2$O$_3$ was obtained without intentionally heating the substrate, either by electron-beam evaporation from Al$_2$O$_3$ pellets or by a strain-induced transition from amorphous to crystalline Al-oxide.

- The correlation of the TLS contribution $\kappa$ to the dielectric susceptibility and structural details shows that O$_2$ in grain boundaries of nanocrystalline γ-Al$_2$O$_3$ is one possible origin of TLS with a high dipole moment and thus a cause for $1/f$ noise in superconducting quantum circuits. Hence, only single crystalline epitaxial Al$_2$O$_3$ tunnel barriers are beneficial to reduce the TLS density.

**Methods**

**Fabrication of Al/AlO$_x$/Al-layer systems**

All Al/AlO$_x$/Al-layer systems were fabricated on single crystalline Si(001)-substrates. The substrates were cleaned with N-ethyl-2-pyrrolidon (NEP), isopropyl alcohol and water to remove the protective resist. All substrates are covered with a 3-4 nm thick native SiO$_x$-layer which forms even after short exposure to air.

The Al/AlO$_x$/Al-structures were fabricated in three different deposition systems. A *MEB 550S (PLASSYS Bestek, Marolles-en-Hurepoix, FR)* electron-beam evaporation system was used for the samples denoted by EBPlas and EBPlas-UV with a base pressure of 10$^{-7}$ mbar. The whole fabrication procedure was carried out without cooling or heating the substrate holder. After the deposition of the bottom Al-layer with a deposition rate of 0.2 nm/s (measured by a piezoelectric sensor) at a deposition pressure of 10$^{-7}$ mbar, the surface was dynamically oxidized by supplying O$_2$ with 10 sccm for 12.5 min. The deposition rate for the top Al-layer was also 0.2 nm/s. The second sample, EBPlas-UV, was fabricated with a slightly increased O$_2$-flux of 12.7 sccm with the same oxidation time of 12.5 min, combined with additional UV-illumination. To remove contamination caused by the bakeout of the UV-lamp, the Si-substrate was plasma-cleaned for 6 min prior to the deposition of the bottom Al-layer. However, the plasma cannot remove the 3-4 nm of SiO$_x$-layer covering the substrate surface. AlO$_x$-layers with thicknesses of ~ 20 nm for dielectric measurements were obtained by repeated deposition of 1 nm Al and oxidation for up to 10 times.

For comparison, an Al/AlO$_x$/Al-layer system, denoted by EBLes, was fabricated in a different electron-beam evaporation system (*PVD 75, Kurt J. Lesker Company, Hastings, UK*). The Al-layers were deposited with a low rate of 0.13 nm/s at a pressure between 1.0 and $1.5 \cdot 10^{-6}$ mbar using a BN-TiB$_2$ crucible with Al-pellets. The AlO$_x$-layer was also deposited by electron-beam evaporation using a second crucible with Al$_2$O$_3$-pellets and a deposition rate of 0.03-0.04 nm/s at a pressure of $10^{-5}$ mbar.

The sample SPUT was fabricated in a home-built sputter deposition system. The substrate was plasma-cleaned with an Ar-flux of 14 sccm at 20 W for 2 min. The deposition of the bottom and top Al-layers took place by Ar-sputtering with an Ar-flux of 19 sccm and a rate of 0.6 nm/s (top layer ) and 0.5 nm/s (bottom layer) at $10^{-3}$ mbar. The Al-target was also used for sputter deposition of the Al-oxide layer by using an Ar/O-plasma (9:1 mixture) at 10 sccm in addition to an increased Ar-flux of 33 sccm at a pressure of $1.4 \cdot 10^{-2}$ mbar resulting in a deposition rate of 0.45 nm/s.

**Transmission Electron Microscopy and Electron Energy Loss Spectroscopy**

Transmission electron microscopy (TEM) was performed with a FEI Titan$^3$ 80-300 microscope operated at 300 kV, which is equipped with an aberration corrector in the imaging lens systems and with a Gatan imaging filter Tridiem HR 865 for electron energy loss spectroscopy (EELS). TEM cross-section specimens were prepared by standard mechanical preparation techniques[49] and Ar$^+$-ion milling with ion energies between 0.5 and 3 keV as final preparation step. Surface oxidation of TEM cross-section samples during transfer from Ar$^+$-ion mill to the microscope is negligible because EELS spectra of Al-layers do not show an O-K signal above the noise level. Structure analyses of nanoscale crystalline regions in the AlO$_x$-layers were performed by comparing two-dimensional Fourier transformation (FT) patterns of HRTEM images with simulated diffraction patterns using the *jems* software[50].

EELS in the scanning transmission electron microscopy (STEM) mode was performed with a convergence angle of 16.7 mrad of the electron probe and a spectrometer acceptance angle of 20.3 mrad. Spectra in Fig. 3 were acquired with a dispersion of 0.02 eV/channel for the Al-L$_{2,3}$ edge and 0.05 eV/channel for the O-K edge. For quantitative composition analysis the Al-$L_{2,3}$ and O-K should be acquired within a single spectrum with high energy resolution. This is not possible because a dispersion of 0.5 eV/channel is necessary to resolve both edges and the count rate of the O-K edge is lower by two orders of magnitude compared to the Al-$L_{2,3}$ edge. Therefore, Al-$L_{2,3}$ and O-K edges were acquired consecutively. To optimize spectrum acquisition for quantitative composition measurement, a self-written script was used which acquires ten to fifty Al-$L_{2,3}$ and O-K spectra alternatingly with a dispersion of 0.05 eV/channel with acquisition times of 0.1 s for the Al-$L_{2,3}$ edge and up to 10 s for the O-K edge combined with binned-gain averaging[51]. The local stoichiometry of the AlO$_x$-layers was quantified on the basis of the k-factor method[52] where γ- and α-Al$_2$O$_3$ with a known composition were used as a reference material. The intensities were measured by signal integration over energy-loss windows with widths between 30 eV and 60 eV taking the different acquisition times for the Al-$L_{2,3}$ edge and O-K edge into account. The variation of the energy-loss integration window width leads to variations of the evaluated composition between 3 - 5 %. Fluctuations of the electron-beam current or microscope alignments contribute to the error, which can be also minimized by alternating acquisition of the two edges. The total error of the chemical composition is estimated to be less than 10 %. We note that the intensity of the sharp peak at 530.7 eV in O-K spectra (Fig. 3c) was not included in the composition quantification because this peak is considered to be an energy loss near edge structure (ELNES) feature that would artificially increase the O-content. Composition data given in Table 1 are based on averaging five to ten measurements at different positions in the center of the AlO$_x$-films.

**Dielectric Measurements**

The Al/AlO$_x$/Al-layer systems of EBPlas, EBPlas-UV and EBLes were deposited to cover the full chip area. Two plate capacitors in series were fabricated for capacity measurements. For this purpose, two areas of each ~1 mm² were covered with photoresist 'Microposit S-1805' (*Dow Chemical Company, Midland, USA*) and wet-etched to remove only the uncovered upper Al-layer using 'Microposit M-319' (*Dow Chemical Company, Midland, USA*). After removal of the upper Al-layer, the etching process is stopped by a dip in distilled water. To determine the etching-time, a test chip for each sample was used where the whole layer-system was removed, because the difference between $AlO_x$-layer and Al-layer cannot be seen during the etching but the Si-substrate is clearly visible. Only 50 % of total etching time was then used for etching our capacitors. As the $AlO_x$ etching-rate is lower by about one order of magnitude, after half of the total etching time, the upper Al-layer is completely removed while the lower Al-layer is still intact. After removal of the photoresist the two remaining areas of the upper Al-layer were contacted by attaching copper wires with silver epoxy EPO-TEK H21 (*Epoxy Technology Inc., Billerica, USA*).

The samples were mounted in a dilution cryostat and capacitance and loss were measured using an Andeen-Hagerling 2500A (*Andeen-Hagerling Inc., Cleveland, USA*) precision capacitance bridge at 1 kHz. For temperature-dependent capacitance measurements, the cryostat is slowly cooled from 1 K to 10 mK within 10-12 h to keep thermal equilibrium with the sample. During this process temperature, capacitance and dielectric loss were measured every 2 min with an acquisition time of 20 s. The temperature was determined by several calibrated temperature-dependent resistors. Using the plate capacitor geometry, the normalized variation of the permittivity can be derived from the capacity according to Eq. (1-3).

**Acknowledgements**

We thank Radian Popescu and Virginia Wilde (Laboratory for Electron Microscopy, Karlsruhe Institute of Technology) for assistance in diffractogram analysis and Martin Obermair (Laboratory for Electron Microscopy, Karlsruhe Institute of Technology) for sample fabrication in the Lesker system. We also acknowledge Saskia Meißner (Physikalisches Institut, Karlsruhe Institute of Technology) for sample fabrication of SPUT and fruitful discussions.

**Author Contributions**

The experiments were conceived by D.G., G.W., A.S. and S.F. and performed by S.F. (TEM measurements) and A.S. (capacitance measurements). Samples were fabricated by L.R. and S.F. The main manuscript text was written by D.G. and S.F. All the authors contributed to data analysis and reviewed the manuscript.

**Competing financial interests**

The authors declare no competing financial interests.

# References

1. Castelvecchi, D. Quantum computers ready to leap out of the lab in 2017. *Nature* **541,** 9–10 (2017).
2. Devoret, M. H. & Schoelkopf, R. J. Superconducting Circuits for Quantum Information: An Outlook. *Science* **339,** 1169–1174 (2013).
3. Sendelbach, S. *et al.* Magnetism in SQUIDs at millikelvin temperatures. *Phys. Rev. Lett.* **100,** 227006 (2008).
4. van Harlingen, D. J. *et al.* Decoherence in Josephson-junction qubits due to critical-current fluctuations. *Phys. Rev. B* **70** (2004).
5. Choi, S., Lee, D.-H., Louie, S. G. & Clarke, J. Localization of metal-induced gap states at the metal-insulator interface: origin of flux noise in SQUIDs and superconducting qubits. *Phys. Rev. Lett.* **103,** 197001 (2009).
6. Holder, A. M., Osborn, K. D., Lobb, C. J. & Musgrave, C. B. Bulk and surface tunneling hydrogen defects in alumina. *Phys. Rev. Lett.* **111,** 65901 (2013).
7. Grabovskij, G. J., Peichl, T., Lisenfeld, J., Weiss, G. & Ustinov, A. V. Strain tuning of individual atomic tunneling systems detected by a superconducting qubit. *Science* **338,** 232–234 (2012).
8. Lisenfeld, J. *et al.* Observation of directly interacting coherent two-level systems in an amorphous material. *Nature communications* **6,** 6182 (2015).

9. DuBois, T. C., Per, M. C., Russo, S. P. & Cole, J. H. Delocalized oxygen as the origin of two-level defects in Josephson junctions. *Phys. Rev. Lett.* **110,** 77002 (2013).

10. DuBois, T. C., Russo, S. P. & Cole, J. H. Atomic delocalization as a microscopic origin of two-level defects in Josephson junctions. *New J. Phys.* **17,** 23017 (2015).

11. Anderson, P. w., Halperin, B. I. & Varma, c. M. Anomalous low-temperature thermal properties of glasses and spin glasses. *Philosophical Magazine* **25,** 1–9 (1972).

12. Phillips, W. A. Tunneling states in amorphous solids. *J Low Temp Phys* **7,** 351–360 (1972).

13. Phillips, W. A. Two-level states in glasses. *Reports on Progress in Physics* **50,** 1657–1708 (1987).

14. Schickfus, M. von & Hunklinger, S. The dielectric coupling of low-energy excitations in vitreous silica to electromagnetic waves. *J. Phys. C: Solid State Phys.* **9,** L439-L442 (1976).

15. Martinis, J. *et al.* Decoherence in Josephson Qubits from Dielectric Loss. *Phys. Rev. Lett.* **95** (2005).

16. Gurvitch, M., Washington, M. A. & Huggins, H. A. High quality refractory Josephson tunnel junctions utilizing thin aluminum layers. *Appl. Phys. Lett.* **42,** 472–474 (1983).

17. Kuiper, A. E. T. *et al.* Plasma oxidation of thin aluminum layers for magnetic spin-tunnel junctions. *J. Appl. Phys.* **89,** 1965 (2001).

18. Krueger, W.H. & Pollack, S. R. The initial oxidation of aluminum thin films at room temperature. *Surface Science* **30,** 263–279 (1972).

19. Rippard, W. H., Perrella, A. C., Albert, F. J. & Buhrman, R. A. Ultrathin aluminum oxide tunnel barriers. *Phys. Rev. Lett.* **88,** 46805 (2002).

20. Tan, E., Mather, P. G., Perrella, A. C., Read, J. C. & Buhrman, R. A. Oxygen stoichiometry and instability in aluminum oxide tunnel barrier layers. *Phys. Rev. B* **71** (2005).

21. Mallison, W. H., Miller, R. E. & Kleinsasser, A. W. Effect of growth conditions on the electrical properties of Nb/Al-oxide/Nb tunnel junctions. *IEEE Trans. Appl. Supercond.* **5,** 2330–2333 (1995).

22. Plisch, M. J., Chang, J. L., Silcox, J. & Buhrman, R. A. Atomic-scale characterization of a Co/$AlO_x$/Co magnetic tunnel junction by scanning transmission electron microscopy. *Appl. Phys. Lett.* **79,** 391 (2001).

23. Zeng, L. J., Krantz, P., Nik, S., Delsing, P. & Olsson, E. The atomic details of the interfacial interaction between the bottom electrode of Al/$AlO_x$/Al Josephson junctions and HF-treated Si substrates. *J. Appl. Phys.* **117,** 163915 (2015).

24. Zeng, L. J. *et al.* Direct observation of the thickness distribution of ultra thin AlOx barriers in Al/AlOx /Al Josephson junctions. *J. Phys. D: Appl. Phys.* **48,** 395308 (2015).

25. Roddatis, V. V. *et al.* The morphology of Al-based submicron Josephson junction. *J. Appl. Phys.* **110,** 123903 (2011).

26. Zeng, L., Tran, D. T., Tai, C.-W., Svensson, G. & Olsson, E. Atomic structure and oxygen deficiency of the ultrathin aluminium oxide barrier in Al/$AlO_x$/Al Josephson junctions. *Sci. Rep.* **6,** 29679 (2016).

27. Zhao, S. P., Finkbeiner, F., Lerch, P., Zehnder, A. & Ott, H. R. Investigations on Nb/Al-AlO$_x$-Al/Nb proximity tunneling junctions for X-ray detection. *J Low Temp Phys* **93,** 641–645 (1993).

28. Schulze, H., Behr, R., Müller, F. & Niemeyer, J. Nb/Al/AlO$_x$/AlO$_x$/Al/Nb Josephson junctions for programmable voltage standards. *Appl. Phys. Lett.* **73,** 996–998 (1998).

29. Clarke, J. & Braginski, A. I. *The SQUID Handbook. Vol. 1: Fundamentals and Technology of SQUIDS and SQUID Systems* (Wiley-VCH, Weinheim, Cambridge, 2002).

30. Otte, H. M. Lattice Parameter Determinations with an X-Ray Spectrogoniometer by the Debye-Scherrer Method and the Effect of Specimen Condition. *J. Appl. Phys.* **32,** 1536–1546 (1961).

31. Zhou, R. S. & Snyder, R. L. Structures and transformation mechanisms of the η, γ and θ transition aluminas. *Acta Crystallogr B Struct Sci* **47,** 617–630 (1991).

32. Bruley, J., Tseng, M.-W. & Williams, D. B. Spectrum-Line Profile Analysis of a Magnesium Aluminate Spinel Sapphire Interface. *Microsc. Microanal. Microstruct.* **6,** 1–18 (1995).

33. Bouchet, D. & Colliex, C. Experimental study of ELNES at grain boundaries in alumina. Intergranular radiation damage effects on Al-L$_{2,3}$ and O-K edges. *Ultramicroscopy* **96,** 139–152 (2003).

34. Balzarotti, A. *et al.* Core Transitions from the Al 2p Level in Amorphous and Crystalline Al$_2$O$_3$. *phys. stat. sol. (b)* **63,** 77–87 (1974).

35. Glans, P. *et al.* Resonant X-Ray Emission Spectroscopy of Molecular Oxygen. *Phys. Rev. Lett.* **76,** 2448–2451 (1996).

36. Hitchcock, A. P. & Brion, C. E. K-shell excitation spectra of CO, N$_2$ and O$_2$. *Journal of Electron Spectroscopy and Related Phenomena* **18,** 1–21 (1980).

37. Kosugi, N., Shigemasa, E. & Yagishita A. High-resolution and symmetry-resolved oxygen K-edge spectra of O$_2$. *Chem. Phys. Lett.* **190,** 481–488 (1992).

38. Enss, C. & Hunklinger, S. *Low-temperature physics* (Springer, Berlin, New York, 2005).

39. Mullins, W.W. The effect of thermal grooving on grain boundary motion. *Acta Metallurgica* **6,** 414–427 (1958).

40. Nik, S. *et al.* Correlation between Al grain size, grain boundary grooves and local variations in oxide barrier thickness of Al/AlO$_x$/Al tunnel junctions by transmission electron microscopy. *SpringerPlus* **5,** 1067 (2016).

41. Kotenev, V. A. Combining Ellipsometry and Electron Microscopy when Identifying the Initial Stages of Aluminum Oxidation. *Protection of Metals* **39,** 415–423 (2003).

42. Thompson, C. grain growth in thin films. *Annual Review of Materials Research* **20,** 245–268 (1990).

43. Verkerk, M. J. & van der Kolk, G. J. Effects of oxygen on the growth of vapor-deposited aluminium films. *Journal of Vacuum Science & Technology A: Vacuum, Surfaces, and Films* **4,** 3101–3105 (1986).

44. Tsuchiya, M., Sankaranarayanan, S. K.R.S. & Ramanathan, S. Photon-assisted oxidation and oxide thin film synthesis: A review. *Progress in Materials Science* **54,** 981–1057 (2009).

45. Jeurgens, L.P.H., Sloof, W.G., Tichelaar, F.D. & Mittemeijer, E.J. Thermodynamic stability of amorphous oxide films on metals: Application to aluminum oxide films on aluminum substrates. *Phys. Rev. B* **62,** 4707–4719 (2000).

46. Oh, S. *et al.* Elimination of two level fluctuators in superconducting quantum bits by an epitaxial tunnel barrier. *Phys. Rev. B* **74** (2006).

47. Weides, M. P. *et al.* Coherence in a transmon qubit with epitaxial tunnel junctions. *Appl. Phys. Lett.* **99,** 262502 (2011).

48. Kumar, P. *et al.* Origin and Reduction of 1/f Magnetic Flux Noise in Superconducting Devices. *Phys. Rev. Applied* **6** (2016).

49. Strecker, A., Salzberger, U. & Mayer, J. specimen preparation for transmission electron microscopy(TEM)-reliable method for cross sections and brittle materials. *Prakt. Metallogr.* **30,** 482–495 (1993).

50. Stadelmann, P. Image analysis and simulation software in transmission electron microscopy. *Microsc. Microanal.* **9,** 60–61 (2003).

51. Bosman, M. & Keast, V. J. Optimizing EELS acquisition. *Ultramicroscopy* **108,** 837–846 (2008).

52. Malis, T. & Titchmarsh, J. M. 'k-factor' approach to EELS analysis. *Electron Microscopy and Analysis* **1986,** 181–184.